\documentclass[doublecol]{epl2} 

\usepackage{bm,amsmath,amssymb,amsfonts}
\usepackage{graphics}

\title{Complete absence of localization in a family of disordered lattices}

\author{Biplab Pal\inst{1}\thanks{E-mail: biplabpal@klyuniv.ac.in} 
\and Santanu K. Maiti\inst{2}\thanks{E-mail: santanu.maiti@isical.ac.in} 
\and Arunava Chakrabarti\inst{1}\thanks{E-mail: arunava\_chakrabarti@yahoo.co.in}}
\shortauthor{Biplab Pal, Santanu K. Maiti and Arunava Chakrabarti}

\institute{                    
  \inst{1} Department of Physics, University of Kalyani, Kalyani, West Bengal-741 235, India\\
  \inst{2} Physics and Applied Mathematics Unit, Indian Statistical Institute, 
  203 Barrackpore Trunk Road, Kolkata-700 108, India
  }

\abstract{We present analytically exact results to show that, 
certain quasi one-dimensional lattices where the building blocks are 
arranged in a random fashion, can have an {\it absolutely continuous} part
in the energy spectrum when special correlations are introduced among 
some of the parameters describing the corresponding Hamiltonians. We 
explicitly work out two prototype cases, one being a disordered array 
of a simple diamond network and isolated dots, and the other an array 
of triangular plaquettes and dots. In the latter case, a magnetic flux 
threading each plaquette plays a crucial role in converting the energy 
spectrum into an absolutely continuous one. A flux controlled enhancement 
in the electronic transport is an interesting observation in the triangle-dot 
system that may be useful while considering prospective devices. The 
analytical findings are comprehensively supported by extensive numerical 
calculations of the density of states and transmission coefficient in 
each case.}

\begin{document}

\maketitle

Localization of electronic states in a disordered lattice, first proposed 
by Anderson~\cite{anderson} is a problem of everlasting interest in 
condensed matter physics, and continues to generate intriguing features 
in quantum transport properties of randomly disordered systems. Over the 
years, with the improvement of fabrication and lithographic techniques the 
realm of Anderson localization~\cite{kramer,abrahams} has extended beyond 
the electronic systems, and has encompassed tailor made artificial crystal 
structures, viz., the photonic~\cite{yablo,john},  
phononic~\cite{montero,vasseur}, plasmonic~\cite{tao,christ} or 
polaritonic~\cite{barinov,grochol}
lattices. Very recently, ultra-cold gases even allowed for the direct 
observation of localization of matter waves~\cite{damski,billy,roati}.

The pivotal result in this field is that, the electronic wave functions 
are known to be localized for dimensions $d \le 2$ (the band center in the 
off diagonal disorder case is an exception), and even for $d > 2$ for 
strong disorder, with an exponential decay in the envelope of the wave 
functions~\cite{kramer,abrahams}. The result has been substantiated 
by meticulous analyses of various calculations related to the 
localization length~\cite{rudo1,rudo2}, density of states~\cite{alberto},
or multi-fractality of the single particles states~\cite{rudo3,rudo4}. 
Extensive work has also been undertaken to study the intricacies of the 
single parameter scaling hypothesis - its validity~\cite{rudo5}, 
variance~\cite{deych}, or even violation~\cite{bunde,titov} in low 
dimensional systems within a tight-binding approximation, that has been 
subsequently consolidated by experimental measurements of conductance 
distribution in quasi-one dimensional gold wires~\cite{mohanty}.

In the last decade we have come across examples, within a tight-binding 
description, where localization-delocalization transitions have been 
observed in disordered systems. The transitions are attributed to certain 
special kinds of correlation in the potential profiles~\cite{dunlap,moura,
maiti} and, have unraveled the presence of discrete energy levels 
corresponding to {\it extended eigenfunctions}~\cite{dunlap}. Experiments 
in this direction~\cite{bellani,khul} have substantiated the theoretical works. 
This led to the possibility of a spectral continuum and metal-insulator 
transition~\cite{moura,maiti} in one, or quasi-one dimensional discrete 
systems. The idea of engineering extended states in two dimensional 
disordered systems~\cite{rudo6} has also been put forward recently. 
In such cases the general exponentially localized character of the eigenfunctions 
prevail, and there is a mixed spectrum of localized and extended states 
(under some special correlations as discussed above).

In the present letter, we present examples where even in a 
disordered arrangement of scatterers, delocalization of electronic states 
can occur throughout the energy spectrum, making it an {\it absolutely 
continuous} one. This happens for a class of quasi one-dimensional systems 
when special correlations are introduced among the numerical values of the 
parameters in the Hamiltonian describing them. Two distinct cases are 
presented. In the first case, we consider a diamond network of four atomic 
like sites placed at random in a host lattice of isolated `atoms' (which 
may be thought of as single level quantum dots). The system is described by 
a tight-binding Hamiltonian. The on-site potential at every lattice point is taken to 
\begin{figure}[ht]
{\centering \resizebox*{8.15cm}{2.6cm}{\includegraphics{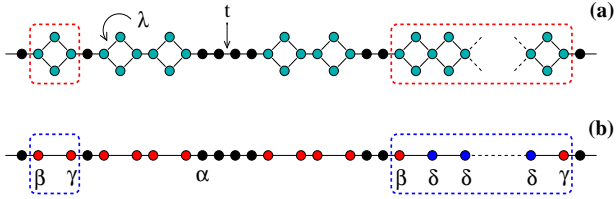}}\par}
\caption{(Color online). (a) A typical realization of an array of isolated 
atomic sites (marked as black circles) and the diamond shaped quadruplets 
(in isolation, and in clusters, and marked by light blue circles). (b) The 
renormalized $1$-$d$ chain, where the isolated dots remain un-renormalized, 
while the diamond clusters are renormalized into $\beta$-$\gamma$ doublets, 
or a sequence of $\beta$-$\delta^n$-$\gamma$. The hopping integral along the 
major axis (backbone) is $t$, while it is $\lambda$ along the edges of a 
diamond. All on-site potentials are equal.}
\label{lattice1}
\end{figure}
be a constant, while the nearest-neighbor hopping integral along the arms 
of the diamond ($\lambda$) differ from that along the back bone ($t$), where 
it is assumed to be constant (see fig.~\ref{lattice1}). There is no 
positional correlation, short range or long range in the conventional 
sense~\cite{dunlap,moura}. We show that, for a particular relationship 
between the nearest-neighbor hopping integrals, the infinite 
{\it diamond-dot} (DD) chain will yield an absolutely continuous spectrum, 
and that, the two terminal transmission coefficient will be unity, 
irrespective of the energy of the electron. 

In the second case, 
the diamond network is replaced by a triangular one leading to an infinite 
{\it triangle-dot} (TD) chain. Each triangular plaquette is threaded by an 
external magnetic flux. The parameters of the Hamiltonian of this TD chain 
are same on that in the DD array. It will be shown that in this case, apart 
from the resonance condition satisfied by $\lambda$ and $t$ as in the DD array, 
the external flux needs to be fixed at a special value to make the spectrum 
absolutely continuous. Thus, in this example, the general localized character 
of the spectrum and the low electronic transmission in an otherwise disordered 
arrangement of scatterers can be enhanced by an external magnetic field. 

In this context, it should be mentioned that the question of inducing 
a whole continuum of extended states in a disordered array of potentials 
has previously been addressed by Rodriguez and Cervero~\cite{rod} using 
a continuous version of the Schr\"{o}dinger equation and with a special 
kind of potential profile. In our case, we focus on a discrete lattice 
model, and the disorder is introduced in the geometrical arrangement 
of the atomic sites along a major axis. The coordination numbers of the 
lattice points range between $2$, $3$ and $4$ in every case, and are 
distributed in a completely uncorrelated, random fashion. This aspect, 
to the best of our knowledge, has not been discussed in the literature 
before. We now discuss the two cases separately.

\textbf{The diamond-dot (DD) system.} -- 
Spinless, non-interacting electrons on the system are described by the 
Hamiltonian,  
\begin{equation}
H_{S}  = \epsilon \sum_{i} d_{i}^{\dagger} d_{i} + \sum_{\langle ij 
\rangle} t_{ij} \left[d_{i}^{\dagger} d_{j} + h.c. \right] 
\label{hamiltonian}
\end{equation}
where, $\epsilon$ is the constant on-site potential, the nearest-neighbor 
hopping integral $t_{ij} = t$ along the backbone, and $t_{ij} = \lambda$ 
along an edge of the diamond. $d_{i}^{\dagger}(d_{i})$ represents the 
creation (annihilation) operator. The Schr\"{o}dinger equation, written 
equivalently in the form of the difference equation, 
$(E - \epsilon) \psi_i = \sum_{j} t_{ij} \psi_{j}$ allows us to decimate 
out the vertices of the diamond networks to map the original chain on to 
an effective one dimensional chain
[fig.~\ref{lattice1}(b)] of four sites $\alpha$, $\beta$, $\gamma$ and 
$\delta$ with on-site potentials 
\begin{align}
&\epsilon_{\alpha} = \epsilon \nonumber \\
&\epsilon_{\beta} = \epsilon_{\gamma} = \epsilon + 2 \lambda^2/(E-\epsilon) \nonumber \\
&\epsilon_{\delta} = \epsilon + 4 \lambda^2/(E-\epsilon).
\end{align}
There is a binary distribution of hopping integrals along the chain now. 
These are the original $t$, and the effective coupling 
\begin{equation}
\tau = 2 \lambda^2/(E-\epsilon)
\end{equation}
arising out of the renormalization of the diamond network. 

Using the difference equation the amplitudes of the wave function at the 
neighboring sites along the effective one dimensional chain can be related 
by the $2 \times 2$ transfer matrices, 
\begin{eqnarray}
\left(\begin{array}{c}
\psi_{n+1} \\
\psi_{n}  
\end{array} \right)
& = & 
\left(\begin{array}{cccc}
\dfrac{E-\epsilon_n}{t_{n,n+1}} & -\dfrac{t_{n,n-1}}{t_{n,n+1}} \\ 
1 & 0 
\end{array} 
\right)
\left(\begin{array}{c}
\psi_{n} \\
\psi_{n-1} 
\end{array} \right)
\end{eqnarray}
A look at the fig.~\ref{lattice1}(b) will make it obvious that there are 
four kinds of transfer matrices, viz, $M_{\alpha}$, $M_\beta$, $M_\gamma$ and 
$M_\delta$, which will differ in their matrix elements, depending on the 
respective on-site potentials and the nearest-neighbor hopping integrals. 
From the arrangement of the diamonds and the isolated sites in the original 
DD chain it can be appreciated that the the wave function at a far end of 
the chain can be determined if one evaluates the product of the unimodular 
matrices $M_{\alpha}$, $M_{\gamma \beta} = M_{\gamma}.M_{\beta}$ and 
$M_{\gamma\delta\beta} = M_{\gamma}.M_{\delta}.M_{\beta}$, or 
$M_{\gamma\delta^{n}\beta} = M_{\gamma}.M_{\delta}^{n}.M_{\beta}$ sequenced 
in the desired random fashion.

The central result of this communication is that, the commutators 
$[M_{\alpha},M_{\gamma\beta}] = 0$, $[M_{\alpha},M_{\gamma\delta\beta}] = 0$, 
and $[M_{\gamma\beta},M_{\gamma\delta\beta}] = 0$ irrespective of 
the energy $E$ of the electron whenever we choose $\lambda = t/\sqrt{2}$. 
To see this explicitly we display the diagonal and the off-diagonal 
elements of the above three commutators. These are,  
\begin{align}
&\left[M_{\alpha},M_{\gamma\beta}\right]_{11}=
\left[M_{\alpha},M_{\gamma\beta}\right]_{22}=0\nonumber\\
&\left[M_{\alpha},M_{\gamma\beta}\right]_{12}=
\left[M_{\alpha},M_{\gamma\beta}\right]_{21}=
\dfrac{(E-\epsilon)(t^2-2\lambda^2)}{2t\lambda^2}
\label{comm1}
\end{align}
\begin{align}
&\left[M_{\gamma\beta},M_{\gamma\delta\beta}\right]_{11}=
\left[M_{\gamma\beta},M_{\gamma\delta\beta}\right]_{22}=0\nonumber\\
&\left[M_{\gamma\beta},M_{\gamma\delta\beta}\right]_{12}=
\left[M_{\gamma\beta},M_{\gamma\delta\beta}\right]_{21}=
\dfrac{(E-\epsilon)(t^2-2\lambda^2)}{2t\lambda^2}
\label{comm2}
\end{align}
\begin{align}
&\left[M_{\alpha},M_{\gamma\delta\beta}\right]_{11}=
\left[M_{\alpha},M_{\gamma\delta\beta}\right]_{22}=0\nonumber\\
&\left[M_{\alpha},M_{\gamma\delta\beta}\right]_{12}=
\left[M_{\alpha},M_{\gamma\delta\beta}\right]_{21}=
\dfrac{(E-\epsilon)(t^2-2\lambda^2)\mathcal{F}}{4t\lambda^4}
\label{comm3}
\end{align}
where, $\mathcal{F}\equiv\mathcal{F}(E,\epsilon,\lambda)=[(E-\epsilon)^2-4\lambda^2]$.

An interesting observation is that, in an infinite 
array of diamonds and dots, without any restriction on randomness, there 
can be isolated diamonds (equivalent to a $\beta$-$\gamma$ pair), or a 
sequence of $n+1$ diamonds (equivalent to a sequence of clusters 
$\beta$-$\delta^{n}$-$\gamma$). The total transfer matrix for the latter 
cluster can be written as, $M_{\gamma\delta^{n}\beta} = U_{n-1}(x) 
M_{\gamma\delta\beta} - U_{n-2}(x)M_{\gamma\beta}$, where, $U_n(x)$ is the 
$n$-th order Chebyshev polynomial of the second kind and 
$x = Tr(M_\delta)/2$. It immediately becomes obvious that 
$[M_{\alpha},M_{\gamma\delta^n\beta}] = 0$ for any value of $n$, as 
$M_{\alpha}$ commutes with both $M_{\gamma\beta}$ and $M_{\gamma\delta\beta}$. 
In addition to this, the commutation of $M_{\gamma\beta}$ and 
$M_{\gamma\delta^{n}\beta}$ also becomes obvious.
This implies that, with $\lambda = t/\sqrt{2}$ the amplitude of the wave 
function or its phase at any lattice point on the {\it renormalized} chain  
[fig.~\ref{lattice1}(b)] in the actual randomly disordered array of diamond 
and dots will be indistinguishable from that {\it in a perfectly periodic 
arrangement of these clusters}. This happens independent of the energy $E$ 
of the electron, that is, throughout the energy spectrum, and is true 
for any kind of disordered arrangement that one can build using a dot, 
isolated diamonds and an array  of any $n$ number of diamonds in juxtaposition. 
The wave functions as result, will have to be of a {\it perfectly extended}, 
Bloch-like character.

\begin{figure}[ht]
{\centering \resizebox*{7.75cm}{1.8cm}{\includegraphics{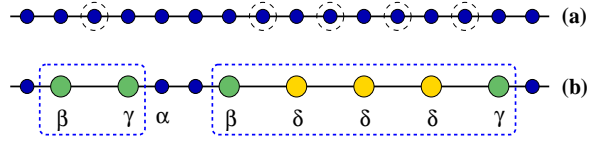}}\par}
\caption{(Color online). (a) A perfectly periodic array of identical atomic 
sites (dots). (b) The renormalized lattice is obtained by decimating the 
encircled dots in (a). This generates a chain where the $\beta$, $\gamma$ 
and $\delta$ sites and the nearest-neighbor hopping integrals have 
precisely those values as obtained in the disordered DD array by setting 
$\lambda = t/\sqrt{2}$.}
\label{order}
\end{figure}

The energy spectrum in the above case is {\it absolutely 
continuous} within the range $[\epsilon-2t,\epsilon+2t]$, and the 
local density of states (LDOS) at any nodal point $\alpha$, $\beta$, $\gamma$ 
or $\delta$ on the renormalized lattice resembles that of a perfectly ordered 
lattice of identical sites with on-site potential $\epsilon$ and nearest 
neighbor hopping integral $t$. We have extensively 
verified this, though present only the average density of states to save 
space. Nevertheless, the fact that it should be the reality, can be tested 
by observing that under the condition $\lambda = t/\sqrt{2}$, the on-site 
potentials on the renormalized chain in fig.~\ref{lattice1}(b) are, 
$\epsilon_\alpha = \epsilon$, $\epsilon_\beta = \epsilon_\gamma = 
\epsilon + t^2/(E-\epsilon)$, and $\epsilon_\delta = \epsilon + 
2 t^2/(E-\epsilon)$. The nearest-neighbor hopping matrix elements turn 
out to be $t$ (unchanged value) and $\tau = t^2/(E-\epsilon)$. Most 
interestingly, identical values can be obtained by beginning with a
perfectly periodic chain with constant on-site potential $\epsilon$ and 
nearest-neighbor hopping $t$, and by arbitrarily decimating sites so as to 
reproduce the same disordered pattern as the original chain. The process 
is illustrated in fig.~\ref{order}. 
Now, the lattice in fig.~\ref{order} 
is a perfectly periodic one, and no matter how we decimate sites, the LDOS 
at any site will be that of a periodic chain with van Hove singularities 
marking the band edges. This is what makes the LDOS at any site on the 
backbone of our original system in fig.~\ref{lattice1}(b) indistinguishable 
from that of an ordered chain of atoms. 

It is also important to look at 
the individual spectra of three subsystems comprising of $\alpha$, $\beta\gamma$ 
and $\beta\delta\gamma$ clusters. The spectrum of a `pure' $\alpha$-lattice 
consists of extended eigenstates spanning an energy interval 
$[\epsilon-2t,\epsilon+2t]$. If we start out with an ordered array of 
$\beta\gamma$ clusters, it is simple to show that the LDOS,   
say, at the $\beta$ site is, 
\begin{equation}
\rho_{\beta}^{(\beta\gamma)}=\dfrac{1}{\pi}\dfrac{(E-\epsilon)^2-2\lambda^2}
{\Bigl(16t^2\lambda^4-(E-\epsilon)^2[(E-\epsilon)^2-t^2-4\lambda^2]^2\Bigr)^{1/2}}
\end{equation}
Similarly, in an ordered array of $\beta\delta\gamma$ clusters, the LDOS at the 
$\beta$ site is of the form,
\begin{equation}
\rho_{\beta}^{(\beta\delta\gamma)}=\dfrac{1}{\pi}\dfrac{\mathcal{A}(E,\lambda,\epsilon,t)}
{\mathcal{B}(E,\lambda,\epsilon,t)}
\end{equation}
where, \begin{math}\mathcal{A}(E,\lambda,\epsilon,t)=
E^4-4E^3\epsilon+\epsilon^4-6\epsilon^2\lambda^2+4\lambda^4
+6E^2(\epsilon^2-\lambda^2)-4E(\epsilon^3-3\epsilon\lambda^2)\end{math} and 
\begin{math}\mathcal{B}(E,\lambda,\epsilon,t)=
\Bigl(64t^2\lambda^8-(E-\epsilon)^2[E^4-4E^3\epsilon+\epsilon^4+
E^2(6\epsilon^2-t^2-8\lambda^2)+4\lambda^2(3\lambda^2+t^2)
-\epsilon^2(8\lambda^2+t^2)+2E\epsilon(8\lambda^2+t^2-2\epsilon^2)]^2\Bigr)^{1/2}\end{math}\\
In each case, on substituting $\lambda=t/\sqrt{2}$, the LDOS at the 
$\beta$ site reduces to the form,
\begin{equation}
\rho_{\beta}^{(\beta\gamma)}=\rho_{\beta}^{(\beta\delta\gamma)}=
\dfrac{1}{\pi}\dfrac{1}{\sqrt{4t^2-(E-\epsilon)^2}}
\end{equation}
which is exactly the LDOS at any $\alpha$ site of a pure, infinite $\alpha$-chain.
Thus, the resonance condition, apart from making the matrices commute, ensures that 
the bands of the respective sub-systems overlap at least within the energy interval 
$[\epsilon-2t,\epsilon+2t]$. In fact, this spectral overlap is essential to ensure complete 
transparency in this energy range together with the extendedness of the wave functions.

To end this part of the discussion we would like to mention that resonant (extended) 
eigenstates  arising out of the commutivity of transfer matrices has been addressed 
also in relation to an array of quantum wells~\cite{gomez}, and a distribution of  
potentials of P\"{o}schl-Teller type~\cite{rodriguez} using a continuous version of 
the Schr\"{o}dinger equation. The difference with the present case is that, here one 
can have commutators independent of the energy of the electron.

To obtain the energy spectrum and the transmission probability numerically
\begin{figure}[ht]
{\centering \resizebox*{7.75cm}{7cm}{\includegraphics{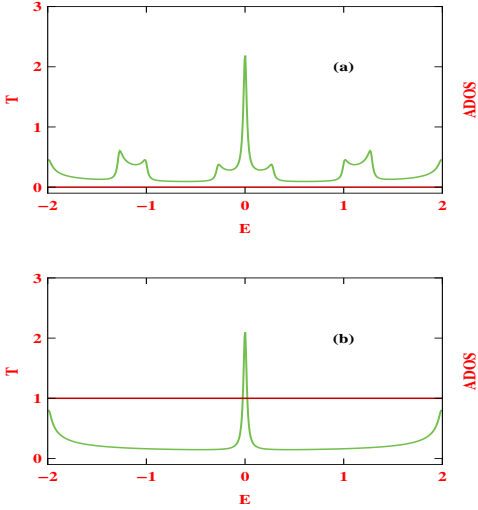}}\par}
\caption{(Color online). Transmission probability $T$ (red color) and 
Average Density of States (ADOS) (green color) as a function of energy 
$E$ for a completely disordered array of $100$ diamonds and $100$ dots.
The results are averaged over $50$ disorder configurations. 
 In the non-resonant case (a) we have chosen $\lambda = 0.3$, 
while $\lambda = 1/\sqrt{2}$ in (b) where we have complete transparency. 
Other parameters are, $t = 1 = t_0$, $\epsilon_i = 0$ at every site, 
$\tau_L = \tau_R = 1$, and $\epsilon_0 = 0$ in the leads.} 
\label{disorderconcen}
\end{figure}
we use a Green's function formalism~\cite{supriya}. Keeping in mind a possible 
experimental realization of the system, we clamp a finite sized system 
of $N$-sites between two ideal semi-infinite electrodes (the left and 
the right electrodes) making an electrode-system-electrode (ESE) bridge. 
The electrodes are described by the Hamiltonian, 
\begin{equation}
H_{0} = \sum_i \epsilon_0 c_i^{\dagger} c_i + \sum_{<ij>} 
t_0 \left(c_i^{\dagger} c_j + c_j^{\dagger} c_i \right)
\label{leadham}
\end{equation}
where different parameters correspond to their usual meaning. The couplings 
between the left (L) and the right (R) electrodes and the system (S) are 
given by, $H_{LS} = \tau_L c_0^\dag d_1 + h.c.$, and 
$H_{RS} = \tau_R d_N^\dag c_{N+1} + h.c.$, so that, the full Hamiltonian of 
the ESE bridge is given by, $H = H_{S} + H_{0} + H_{LS} + H_{RS}$. 
The average density of states (ADOS) of the ESE bridge is given by, 
\begin{equation}
\rho_{av} = - \frac{1}{N \pi} \mbox{Im} \left[\mbox{Tr}(G)\right] 
\label{ados}
\end{equation}
where, $G=\left(E-H+ i \eta \right)^{-1}$, with $\eta \rightarrow 0$.  

In terms of the Green's function of the system and its coupling to 
the side-attached electrodes, the transmission probability can be written 
in the form~\cite{supriya},
\begin{equation}
T={\mbox{Tr}} \left[\Gamma_L \, G_{S}^r \, \Gamma_R \, G_{S}^a\right]
\label{trans}
\end{equation}
where, $\Gamma_L$ and $\Gamma_R$ describe the coupling of the system to 
the left and right electrodes, respectively. Here, $G_{S}^r$ and $G_{S}^a$ are 
the retarded and advanced Green's functions, respectively, of the system 
including the effects of the electrodes. 

The full system is  
partitioned into sub-matrices corresponding to the individual sub-systems 
and the Green's function for the disordered sample is effectively 
written as,
\begin{equation}
G_{S}=\left(E-H_{S}-\Sigma_L-\Sigma_R \right)^{-1}
\label{greensfn}
\end{equation}
where, $\Sigma_L$ and $\Sigma_R$ are the self-energies due to coupling of 
the system to the left and right electrodes, respectively~\cite{supriya}. 
All information of the coupling are included in these two self-energies.
\begin{figure}[ht]
{\centering \resizebox*{7.2cm}{4.2cm}{\includegraphics{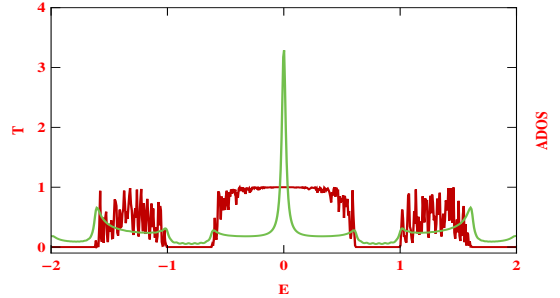}}\par}
\caption{(Color online). Transmission probability $T$ (red color) and ADOS 
(green color) as a function of energy $E$ for the diamond-dot chain away 
from the resonant case. The results averaged over $50$ disorder 
configurations have been presented. $100$ diamonds with $100$ dots have 
been taken with $\lambda = 0.98499$, a forty percent deviation from the 
resonance value of $1/\sqrt{2}$. Other parameters are the same as in the 
earlier figures.}
\label{deviation}
\end{figure}

In fig.~\ref{disorderconcen} we show the ADOS and 
the transmission coefficient of a random sequence of $100$ diamonds and 
$100$ dots. Our previous analytical argument for the infinite system is perfectly 
corroborated in these diagrams. Complete localization of all the eigenstates, 
corresponding to zero transmission, is observed [fig.~\ref{disorderconcen}(a)]
as long as $\lambda$ is set quite arbitrarily. The scenario changes 
as soon as the value of $\lambda$ is set equal to the resonant value of 
$1/\sqrt{2}$ (with $t=1$). We now see an absolutely continuous energy 
spectrum between $E = \pm 2t$, the central peak at $E=0$ being 
the contributions from the top and bottom sites of the diamond plaquettes, and 
the transmission coefficient becomes equal to unity {\it irrespective of the energy} 
of the electron.

Keeping in mind an experimental realization of such 
a finite size sample, we have carefully studied the effect of a possible 
deviation in the value of $\lambda$ from its ideal `resonance value' of 
$1/\sqrt{2}$. The continuity of the spectrum stands out to be a robust 
result even when the deviation is substantial. In fig.~\ref{deviation} 
we show the ADOS and the transmission coefficient for a $40\,\%$ deviation 
in the value of $\lambda$ from its resonant value. The transmission spectrum 
is seen to be split into three continuous sub-bands of high transmission. 
The central sub-band still retains its continuous character. 
The reason is, the commutators [Eq.~\eqref{comm1} - \eqref{comm3}] 
vanish whenever $E=\epsilon$, even for an arbitrary value of $\lambda$. Then 
the band center corresponds to an extended state. The transmission coefficient 
will be almost unity for a finite sized system even when we are off-resonance.

\begin{figure}[ht]
{\centering \resizebox*{8.3cm}{2.0cm}{\includegraphics{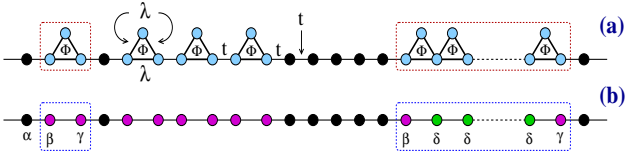}}\par}
\caption{(Color online). (a) A disordered arrangement of triangular 
plaquettes threaded by a magnetic flux $\Phi$, and dots (marked by 
black circles). The inter-atomic hopping integral between the sites of 
the triangular plaquette is $\lambda$, and the hopping integral has a 
value $t$ between two consecutive triangles or dots, or in between a 
triangle and a dot. (b) The effective $1$-d chain obtained by renormalizing 
the lattice in (a).}
\label{lattice2}
\end{figure}

\textbf{The triangle-dot (TD) system.} -- We now move to our second example where, 
we have a disordered array of triangular plaquettes and isolated `dots' [fig.~\ref{lattice2}(a)]. 
Each arm of a triangle is assigned a hopping integral $\lambda$, and every triangle is threaded
by a magnetic flux $\Phi$. The hopping integral in between triangles and dots is 
equal to $t$ and the on-site potential at every lattice point is $\epsilon$ as before.
Once again, decimating the vertices of every triangle we arrive at an effective 
one dimensional chain [fig.~\ref{lattice2}(b)] where time reversal symmetry is 
broken across the `bonds' joining the base atoms (red colored points) of a triangle.
The sites residing at the `left' and the `right' corners of a triangle will be called 
the $\beta$- and $\gamma$-sites with an effective on-site potential given by,
\begin{equation}
\epsilon_\beta = \epsilon_\gamma = \epsilon+\dfrac{\lambda^2}{E-\epsilon}
\end{equation}
The hopping integral connecting a $\beta-\gamma$ pair is now given by,
\begin{equation}
t_F = \dfrac{\lambda^2 e^{2i\theta}}{E-\epsilon} + \lambda e^{-i\theta} 
\end{equation}
when an electron `hops' from a $\beta$ to a $\gamma$ site, and $t_B = t_F^*$ while 
hopping `backward' - a consequence of broken time reversal symmetry. 
In the above equation $\theta=2\pi\Phi/3\Phi_{0}$ where, $\Phi_{0}=hc/e$ is the 
fundamental flux quantum.
 
One can now look for a possible commutation of the same matrices, viz., 
$M_{\alpha}$, $M_{\gamma\beta}$ and $M_{\gamma\delta\beta}$ in the spirit of the DD 
case in earlier section. In every case, the diagonal elements of the commutators 
turn out to be zero, and the off diagonal elements are given by,
\begin{align}
&\left[M_{\alpha},M_{\gamma\beta}\right]_{12} = 
-e^{i\theta} \dfrac{\chi}{\lambda t\Delta} \nonumber \\
&\left[M_{\alpha},M_{\gamma\delta\beta}\right]_{12} = 
-e^{2i\theta} \dfrac{(E-\epsilon)^2 - 2\lambda^2}{\lambda^2t\Delta^2}\chi
\nonumber \\
&\left[M_{\gamma\beta},M_{\gamma\delta\beta}\right]_{12}  = 
-\dfrac{(E-\epsilon) e^{3i\theta} + \lambda}{\lambda t\Delta^2} \chi 
\end{align}
where, $\Delta=E-\epsilon+\lambda \exp{3i\theta}$ and, 
\begin{equation}
\chi = 2\lambda^3\cos 3\theta - (E-\epsilon) (t^2-2\lambda^2)
\label{chieq}
\end{equation}
The $(2,1)$ element of every commutator matrix is equal to its $(1,2)$
element.

A look at Eq.~\eqref{chieq} immediately reveals that, selecting $\Phi=\Phi_{0}/4$ and
$\lambda=t/\sqrt{2}$ makes every commutator vanish. This is the desired result.
The entire randomly disordered array of triangular plaquettes and dots then becomes 
indistinguishable from a perfectly periodic arrangement of the same. In the same spirit, 
as done in the earlier DD case, we have examined the overlap of the LDOS spectrum of 
the lattices comprising of $\alpha$, $\beta\gamma$ and $\beta\delta\gamma$ clusters.
The spectrum of each individual periodic system has a common region of overlap again 
between $E=\epsilon\pm2t$ as soon as one sets $\Phi=\Phi_{0}/4$ and
$\lambda=t/\sqrt{2}$. 
\begin{figure}[ht]
{\centering \resizebox*{7.75cm}{7cm}{\includegraphics{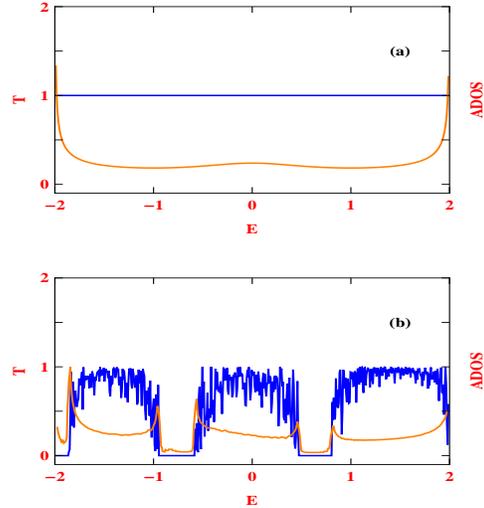}}\par}
\caption{(Color online). Transmission probability $T$ (blue color) and 
ADOS (orange color) as a function of energy $E$ for an array of $150$ 
triangles and $150$ dots. We have chosen $\epsilon=0$, $t=1$, and $\lambda=
1/\sqrt{2}$. (a) $\Phi=0.25\Phi_{0}$, the resonant case, and (b) $\Phi=
0.15\Phi_{0}$, the off-resonance case. The lead parameters are same as in 
the other figures.} 
\label{tdresonance}
\end{figure}
Thus, as before, we expect a continuum of extended states and 
perfect transmission irrespective of energy in this region as long as the resonance 
conditions are satisfied. The important distinction with the DD case is that, now we 
are able to fine-tune the transport in a disordered TD array by dint of an {\it external 
magnetic field} - a fact that might be useful from the standpoint of designing electronic 
devices.

The ADOS and the transmission spectrum in the resonant and off-resonant conditions 
are displayed in fig.~\ref{tdresonance} for a system of $150$ triangles and $150$ 
dots placed randomly along a line. The ballistic character of the transmission 
coefficient in part (a) of the figure is perfectly in accord with the analysis 
presented here. In the off-resonance condition we again have sub-bands of high 
transmittivity. The central band still remains continuous around $E=\epsilon$. 
The reason is again attributed to the fact that at $\Phi=\Phi_{0}/4$ the 
commutivity may also be achieved by setting $E=\epsilon$. This makes the central 
band survive for a finite sized system even when we deviate from the resonance condition.
\acknowledgments
Biplab Pal would like to thank DST, India for providing financial 
assistance through an INSPIRE Fellowship (IF110078). Santanu K. Maiti 
is thankful to Abraham Nitzan for stimulating discussions, and gratefully 
acknowledges the hospitality at the School of Chemistry, Tel Aviv University 
where part of the computation was done.

\end{document}